\documentclass[aps,prc,twocolumn,showpacs,floatfix]{revtex4}
\usepackage{epsf,epsfig}

\newcommand{\be}{\begin{equation}}
\newcommand{\ee}{\end{equation}}
\newcommand{\bea}{\begin{eqnarray}}
\newcommand{\eea}{\end{eqnarray}}

\newcommand{\gton}{\mathrel{\lower.9ex \hbox{$\stackrel{\displaystyle 
>}{\sim}$}}} 
\newcommand{\lton}{\mathrel{\lower.9ex \hbox{$\stackrel{\displaystyle 
<}{\sim}$}}}

\newcommand{\vx}{{\bf x}}

\newcommand{\vp}{{\bf p}}

\begin{document}

\title{High-{\boldmath $p_T$} 
suppression and elliptic flow from radiative energy loss
with realistic bulk medium expansion}

\author{Denes Molnar}
\author{Deke Sun}
\affiliation{Physics Department, Purdue University, West Lafayette, IN 47907}

\date{\today}

\begin{abstract}
We investigate nuclear suppression and elliptic flow
in $A+A$ reactions using Gyulassy-Levai-Vitev (GLV) radiative energy loss
with the covariant transport MPC for bulk medium evolution.
At both RHIC and LHC energies, we find that inclusion of
realistic transverse expansion for the medium strongly suppresses 
elliptic flow at high $p_T$ compared to calculations with longitudinal
Bjorken expansion only. We argue that this is a generic
feature of GLV energy loss.
Transverse expansion also enhances the high-$p_T$ suppression, 
while fluctuations in energy loss with the rescattering location
of the jet parton in the medium lead to weaker
suppression and smaller elliptic flow.
However, unlike the strong 
reduction of elliptic flow with transverse expansion,
these latter effects get nearly washed out once calculations are
adjusted to reproduce $R_{AA}$ in central collisions.
\end{abstract}


%

\maketitle

\section{Introduction}

Understanding parton energy loss in ultrarelativistic heavy-ion reactions has 
been the focus of considerable recent theoretical effort. A variety of
phenomenological approaches
(e.g., \cite{Jia_Eloss,Liao_Shuryak_Eloss,Betz_Eloss}) 
formulate the problem deterministically, in terms of 
a local energy loss rate $dE/dL = - f(E(L),T(L), L)$
that depends on local temperature, position, and parton energy
along the 
Eikonal (straight-line) parton
trajectory.
In the weak-coupling regime, more rigorous treatment is possible based on
perturbative QCD\cite{BDMPS,WS,GLV}.
This includes quantum interference effects and also 
fluctuations, namely, energy loss for a given jet becomes a 
stochastic variable that is a 
function of the scattering and emission history of the jet.

A critical step in computing heavy-ion observables from any energy loss model 
is spatial and temporal averaging over the bulk medium formed in the 
collision. We employ here the Gyulassy-Levai-Vitev (GLV) 
framework\cite{GLV} 
in which a high-energy parton loses energy
through gluon radiation induced by interactions with static Yukawa
scatterers in the medium. 
It was natural for us to combine this approach with parton transport 
for the bulk evolution, namely Molnar's Parton Cascade\cite{MPC} (MPC).
Selected early findings were highlighted in~\cite{HP2012}. 
Here we present a comprehensive set of results 
from a numerically improved
calculation.

Our approach is similar to recent work by
Buzzatti and Gyulassy\cite{CUJET1p0} or Horowitz\cite{Horowitzv2},
but with a few key differences. Unlike \cite{CUJET1p0}, 
we only focus here on light partons, and do not include multiple 
gluon radiation, elastic energy loss, or 
energy loss fluctuations due to variations in radiated gluon 
momentum.
However, as in \cite{HP2012}, 
we {\em do} include medium evolution with realistic 3D expansion,
both longitudinal and 
transverse, which turns out to crucially influence elliptic flow
and also affect the nuclear suppression factor.
We also study how differences between deterministic (average) 
energy loss and stochastic energy loss, which fluctuates
depending on
where the jet parton interacts with the medium, impact observables.

Recent assessment by the PHENIX Collaboration\cite{PHENIXeloss} highlighted
the difficulty perturbative QCD
parton energy loss frameworks have with reproducing
nuclear suppression at RHIC as a function of the angle with the reaction plane,
$R_{AA}(\phi)$. GLV calculations were noticeably absent from that analysis
but our work here provides a very similar cross-check of the GLV framework.
The two observables we study, 
azimuthally averaged (traditional) $R_{AA}$ and elliptic flow $v_2$,
carry essentially the same information%
\footnote{Roughly speaking,
$R_{AA} \approx [R_{AA}(\phi{=}0) + R_{AA}(\phi{=}\pi/2)]/2$,
while $v_2 \approx [R_{AA}(\phi{=}0) - R_{AA}(\phi{=}\pi/2)]/(4R_{AA})$.
}
as $R_{AA}(\phi)$.

\section{Ingredients of the calculation}
\label{Sec:GLV}

\subsection{Energy loss framework}

We consider here the Gyulassy-Levai-Vitev (GLV) 
formulation of energy loss\cite{GLV},
which provides the spectrum of gluons radiated by a jet parton
as an expansion in the total number of scatterings $n$
the jet and the radiated gluon experience with the medium as they travel 
through it.
The dominant contribution is given by the leading $n=1$ (single scattering) 
term
\bea
x\frac{dN^{(1)}}{dx\, d^2 {\bf k}} &=&
\frac{C_R \alpha_s}{\pi^2} \chi 
\int d^2 {\bf q} \, 
\frac{\mu^2(z)}{\pi [{\bf q}^2 + \mu^2(z)]^2}
\, \frac{2 {\bf k}{\bf q}}{{\bf k}^2 ({\bf k} - {\bf q})^2}
\nonumber\\
&&\qquad\qquad\quad\times
\left( 1 -   \cos \frac{({\bf k} - {\bf q})^2 z}{2xE} \right) 
\ .
\label{GLV1}
\eea
Here, $x$ and ${\bf k}$ are  the light-cone momentum fraction and 
transverse momentum of the radiated gluon,
the jet energy is $E$, the original hard scattering is at $z=0$,
the jet is moving in the 
$z$ direction and rescatters once at position $z$, $\mu(z)$ is the local
Debye screening mass, $\chi = \int dz\, \rho\, \sigma$ is the opacity for 
{\em gluon} jets 
(irrespectively of the jet Casimir $C_R$), and
$\sigma = 9\pi \alpha_s^2/(2\mu^2)$ is the 
(screened) total $gg\to gg$ scattering cross section
provided the medium is made of gluons.
The result 
was obtained for a medium of static Yukawa scatterers that are
well-separated with mean distance $d \gg 1/\mu$,
the radiated gluon is assumed to be soft ($x \ll 1$), and the calculation
was done at tree level.
Higher-order terms in number of scatterings 
have been systematically investigated\cite{Wicks_GLV_n}, up to at least $n=9$,
and were found to give modest corrections to the $n=1$ result.

We integrate the spectrum (\ref{GLV1}) numerically 
to obtain a momentum-averaged 
energy loss
\be
\Delta E^{(1)}(z) = \int dx\, d^2{\bf k}\,
E x\, \frac{dN^{(1)}}{dx\, d^2 {\bf k}}
\label{Eq:deltaEz_gen}
\ee
for {\em fixed} $z$,
i.e., retain energy loss fluctuations due to variations in $z$ only.
The probability for the scattering to occur at $z$ is
\be
p(z) = \frac{\rho(z) \sigma(z)}{\chi} \ .
\ee 
Integrating over $z$ as well yields 
the deterministic average energy loss
\be
\langle\Delta E^{(1)}\rangle = \int dz \, p(z)\, \Delta E^{(1)}(z) \ .
\label{Eq:avE}
\ee

\begin{figure}[h]
\epsfysize=6.15cm
\epsfbox{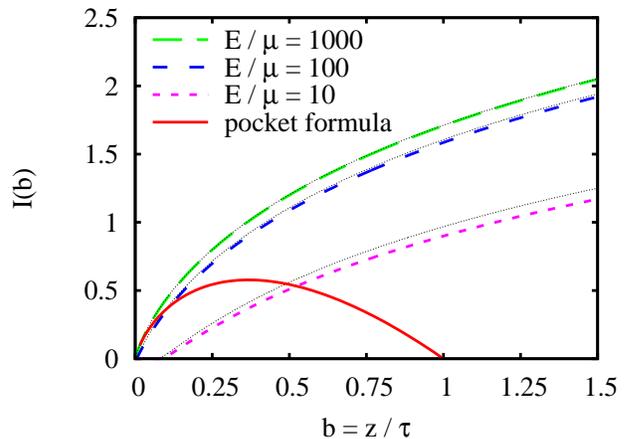}
\caption{Function $I(b,E/\mu)$ that governs GLV energy loss $\Delta E(z)$
in this work (see text). Simple analytic ``pocket formula'' (solid line) valid
for asymptotic jet energies and small $z/\tau$ is compared to the result with
kinematic cutoffs
(dashed lines) for various jet energies.
Thin dotted lines are for an approximate calculation of $I$ 
in our earlier work~\cite{HP2012}.
}
\label{Fig:I}
\end{figure}

For asymptotic jet energies, the average energy loss (\ref{Eq:avE}) is given by
the GLV ``pocket formula'' \cite{GLVv2}
\be
\langle\Delta E^{(1)}\rangle \approx \frac{9\pi C_R \alpha_s^3}{4}
   \int dz\, z\, \rho(z) \ln \frac{2E}{\mu^2 z}
\label{Eq:pocketGLV}
\ee
Though this remarkably compact expression 
(and its even simpler cousin without the $\ln[2E/(\mu^2 z)]$ factor) 
has motivated workers in the past, 
it is unreliable in practice even for qualitative conclusions.
Basic constraints
that i) the radiated gluon energy does not exceed the energy of the parent jet,
ii) momentum transfer in scattering off a thermal particle in the medium 
has {\em on average} an upper bound given by the center-of-mass energy 
available, 
and iii) the radiated gluon energy must exceed the plasma frequency
for the gluon to propagate in the medium, translate into kinematic bounds
\be
 k < x E\ , \quad q < \sqrt{6ET} \ , \quad x E \gton \mu \ ,
\label{kin_bounds}
\ee 
and lead to (see App.~\ref{App:I})
\be
\Delta E^{(1)}(z) = \frac{2C_R \alpha_s}{\pi} \, E\, \chi
\,I\!\left(b=\frac{z}{\tau(z)}, \epsilon = \frac{E}{\mu(z)}\right)
\label{Eq:deltaEz}
\ee
where $\tau(z) = 2 E / \mu^2(z)$ is a characteristic formation time, 
and the function 
$I$ is determined by the integral (\ref{Eq:I}).
For small $b \to 0$ and large $E/\mu\to \infty$,
$I(b,E/\mu\to\infty) \approx - (\pi/2)\, b\, \ln b$, and one recovers
(\ref{Eq:pocketGLV}). However, as shown in Fig.~\ref{Fig:I},
the pocket formula
does not account for smaller energy energy loss for less energetic jets, and
it also shows a peculiar maximum and turnover at higher $z/\tau$.

Jet and medium parameters
enter only through the arguments of $I$, 
which simplifies the calculation considerably because 
we can precalculate a 2D
table and interpolate later.
Tabulating $I$ is still time consuming
because the integrand is oscillatory. To speed up the numerics, 
in~\cite{HP2012}
we used the approximate expression (\ref{Eq:I_2D}) that involves one fewer
integrals. Here we use the
full result
(\ref{Eq:I}) and confirm in Fig.~\ref{Fig:I}
the accuracy of our earlier calculation (thin dotted lines).

For very small $b\approx 0$, $I$ can go slightly negative, 
implying a small
energy gain instead of energy loss.
In the calculation
we cut these off and set $I = 0$ whenever negative
contributions arise
(see, e.g., the curve for $E/\mu = 10$ below
$b < 0.1$ in  Fig.~\ref{Fig:I}).

We emphasize that
here we only consider radiative energy loss as encoded in (\ref{GLV1}).
There are significant additional contributions from 
collisional energy loss\cite{GLV_rad+el}. 
Recoil of scattering centers in scattering also
enhances\cite{DGLV_dyn}
energy loss because it effectively changes
the denominator $({\bf q}^2 + \mu^2)^2$ in (\ref{GLV1}) to
${\bf q}^2({\bf q}^2+\mu^2)$.
One also expects potentially large 
higher-order corrections in the strong coupling constant 
$\alpha_s$ that are yet to be computed.
In our approach all these effects 
get roughly incorporated into the parameter $\alpha_s$ when we calibrate
the calculation to a subset of observables through $\Delta E \sim \alpha_s^3$.
Equivalently, throughout this paper we
apply (\ref{GLV1}) with {\em rescaled} opacities
$\chi \to Z \chi$ with scaling factor
$Z = 3$ (i.e., $\Delta E \sim Z \alpha_s^3$).

\subsection{Coupling to bulk medium}
\label{Sec:coupling}

Coupling energy loss to bulk medium evolution 
involves several open questions and assumptions
(see, e.g., \cite{RenkSystematics} and references therein 
for a detailed discussion).
First, the GLV result (\ref{GLV1}) was derived for a medium
of {\em static} scattering centers whereas in a heavy-ion collision the density
changes rapidly with time. Unfortunately,
the formulation has not been extended yet to 
time-dependent background color fields.
Therefore, as is customary
in practice, for {\em non-static} media we
reinterpret $\rho(z)$ in the GLV formula as the local density 
$\rho(z,t=t_0 + z)$ along the parton trajectory.

Second, it is nontrivial to 
apply perturbative energy loss unless the medium is also 
perturbative. For example, questions arise
whether the scatterers are quark- or gluon-like quasiparticles or 
some other field configurations, and whether 
the key parameter in (\ref{GLV1}), the local Debye mass $\mu \sim gT$,
is computed from the temperature ($\sim T$), 
local energy density ($\sim e^{1/4}$), 
or some other 
thermodynamic quantity. 
For simplicity, 
we will consider here
the medium to be a
system of massless gluons that thermalize to a large degree by Bjorken 
proper time $\tau_0 = 0.6$~fm.
The local $\mu(z)$ is then determined
along the 
straight-line jet path from the local temperature,
which we calculate using  
tabulated density evolution $n(\vx_\perp, \eta = 0, \tau)$
from the bulk dynamics model. For our gluon gas, $n \approx 2 T^3$ 
and $\mu = gT \approx 2 T$ 
(similar expressions have been used in~\cite{GLVv2}). 
We set aside the interesting question of quark chemical 
equilibration,
namely, how quark/antiquark abundances build up after the early
gluon-dominated stage of the collision. 
At fixed parton density, quarks would significantly
lower the opacity because
of their smaller Casimir $C_2$. (On the other hand, the Debye mass
would only be affected by $\sim 10$\% based on perturbation theory%
\footnote{Boldly applying at temperatures $T \sim 150-700$~MeV
accessible at RHIC and the LHC perturbative quark-gluon plasma expressions
for density and Debye mass,
$$
\frac{n}{T^3} = (16 + 9 N_f) \frac{\zeta(3)}{\pi^2}, \ 
\quad \frac{\mu}{gT} = \sqrt{1+\frac{N_f}{6}} \ ,
$$
$\mu(T(n))$ in a pure gluon plasma ($N_f = 0$) is not that different from a
plasma with light quarks ($N_f\approx 2-3$)
because the $N_f$ dependences largely compensate each other.}%
.)

Another open question is energy loss prior to the assumed thermalization time
$\tau_0$.
Reference~\cite{CUJET1p0} found GLV energy loss to be quite sensitive to
whether density for $\tau < \tau_0$ was taken to be constant, or rising
linearly with $\tau$, or zero. However, differences in early density evolution
were largely compensated by slightly different values 
of $\alpha_s$ when the calculation was 
calibrated to experimental data.
We assume here at early times $\tau < \tau_0$ a 
linear density build-up 
$\rho(\vx_\perp, \eta, \tau) = \tau \rho(\vx_\perp, \eta, \tau_0) / \tau_0$.

One also has to address energy loss in the low-temperature (hadronic) 
regime, where the isolated Yukawa scattering center assumption in GLV 
clearly breaks down.
In the calculation here we do not include hadrons, and 
simply set vanishing $\Delta E(z) = 0$ 
for jet path sections where $\mu < \Lambda_{QCD}$, 
with cutoff $\Lambda_{QCD} = 0.2$~GeV.

Though we expect that our qualitative conclusions are robust,
it would be interesting to vary the assumptions above in the future.

\subsection{Jet initial conditions and fragmentation}

We take initial jet momentum distributions in p+p, Au+Au and Pb+Pb
from leading-order (LO) perturbative QCD with one-loop running
coupling $\alpha_s(Q^2)$, using CTEQ5L parton distribution function 
parameterizations\cite{CTEQ5L} with $Q^2 = p_{T,parton}^2$.
Nuclear effects such as shadowing are ignored but isospin
(proton-neutron difference) is included.
We consider jets produced at
midrapidity both in coordinate and momentum space,
i.e., $y = \eta = 0$, where $y$ and $\eta$
are the coordinate and momentum rapidity%
\footnote{With the $z$ axis along the beam direction,
$$
y = \frac{1}{2} \ln \frac{E+p_z}{E-p_z} \ , \qquad
\eta = \frac{1}{2} \ln \frac{t + z }{t - z}
$$
}.
The jet transverse momentum distribution is generated uniformly in azimuth,
while the transverse density distribution follows
the binary collision
distribution for two Woods-Saxon distributions
$\rho_A(r)$. I.e.,
for collisions with impact parameter $b$ we use the transverse profile
\be
\frac{dN({\bf b})}{d^2 {\bf x}_\perp} = const \times
T_A\!\!\left({\bf x}_\perp +\frac{{\bf b}}{2}\right)
T_A\!\!\left({\bf x}_\perp -\frac{{\bf b}}{2}\right)
\; \; ,
\label{Eq:T_AB}
\ee
where $T_A({\bf b})\equiv \int dz \rho_A(\sqrt{z^2+{\bf b}^2})$
is the nuclear thickness function.
Woods-Saxon parameters
for gold and lead nuclei were taken from HIJING\cite{HIJING}.
To enhance statistics at high $p_T$, we generate uniform jet $p_T$
in the intervals $[2,80]$~GeV for RHIC and $[2,300]$~GeV for LHC,
and then appropriately reweight contributions to reflect the real jet
spectrum.

The density evolution of scatterers is calculated as discussed 
in the next Section.
After energy loss, jets are fragmented
independently
using LO BKK95 fragmentation function parameterizations\cite{BKK95} 
with scale factor
$Q^2 = p_{T,hadron}^2$, and we take
$\pi_0 = (\pi^+ + \pi^-) / 2$ for the neutral pion yield.
This procedure
reproduces high-$p_T$ $\pi_0$ and charged particle 
spectra in p+p at RHIC and LHC with a modest $K$-factor 
$K_{NLO} \approx 2.5$ to account for higher-order contributions.

\subsection{Bulk medium evolution for $\tau > \tau_0$}
\label{Sec:transport}

For medium evolution {\em without} transverse expansion, we use the initial
conditions discussed later in this Section but keep
the transverse profile ``frozen'', i.e., as in~\cite{CUJET1p0,Horowitzv2},
the local density undergoes 
longitudinal Bjorken expansion 
$\rho(\vx_\perp,\eta,\tau) = \rho(\vx_\perp,\eta,\tau_0) \tau_0 / \tau$.

For medium evolution {\em with} transverse expansion,
we employ covariant transport theory
in the same spirit as Refs.~\cite{hytrv2,isvstr0}
to model the collective 
expansion of a causal, relativistic, low-viscosity bulk medium.
We  evolve a system of 
massless ``gluons'' with $2\to 2$ interactions via MPC\cite{MPC} 
(Molnar's Parton Cascade), i.e., solve
\bea
p_1^\mu \partial_\mu f_1 &=& S(x, \vp_1) 
+ \int\limits_2\!\!\!\!
\int\limits_3\!\!\!\!
\int\limits_4\!\!
\left(f_3 f_4 - f_1 f_2\right)
W_{12\to 34} \nonumber\\
&&\qquad\qquad\quad\times \  \delta^4(p_1{+}p_2{-}p_3{-}p_4)
 \ .
\label{Eq:BTE}
\eea
Here  the integrals are shorthands
for $\int_i \equiv \int d^3 p_i / (2E_i)$,
and $f_j \equiv f(x, \vp_j)$ is the phase space density.
The source function $S$ specifies the initial conditions,
while $W=(1/\pi)s^2 d\sigma/dt$ controls the local scattering rate.

We use isotropic $d\sigma/dt = 2\sigma_{tot} / s$, for simplicity, with
$\sigma_{tot}(\tau) = \sigma_0 (\tau/\tau_0)^{2/3}$ growing with time 
so that 
the shear viscosity to entropy ratio $\eta/s$ stays approximately
constant\cite{minvisc,isvstr0}. 
To generate substantial elliptic flow 
$v_2(p_T\approx 3\ {\rm GeV}) \sim 0.25$ 
in collisions with $b=8$ fm impact parameter at RHIC and LHC,
we set $\sigma_0=8$~mb and $4.5$~mb respectively.

Initial conditions
for Au+Au at $\sqrt{s_{NN}} = 200$~GeV and 
Pb+Pb at $\sqrt{s_{NN}} = 2.76$~TeV with impact parameter $b=3$ and $8$~fm 
(about 0-10\% and 30\% centrality, respectively) are constructed as follows.
As in~\cite{hytrv2}, we start at proper time $\tau_0 = 0.6$~fm 
with a locally thermalized system at temperature $T=0.385$~GeV.
Because we 
are only interested in observables
at midrapidity, we set up longitudinally boost invariant conditions 
in a wide coordinate rapidity window $|\eta| < 5$, 
with rapidity densities $dN(b)/d\eta \propto N_{part}(b)$ proportional
to the number of participants (``wounded'' nucleons)
\bea
N_{part}(b) &=& \int d^2 {\bf x_\perp} \, 
\left[T_A({\bf x}_\perp + {\bf b}) +T_A({\bf x}_\perp - {\bf b})\right]
\nonumber \\ 
&& \qquad\qquad\quad \times \ 
\left(1 - e^{-\sigma_{NN} T_A({\bf x}_\perp)}\right) \ .
\eea
Here the inelastic 
nucleon-nucleon cross section is $\sigma_{NN} = 42$~mb at $\sqrt{s_{NN}}=200$~GeV (RHIC) and
$70$~mb at $\sqrt{s_{NN}}=2.76$~TeV (LHC),
and we match the ``gluon'' multiplicity 
$dN/d\eta$ to the observed charge particle $dN/dy$
via setting $dN(b=0)/d\eta = 1100$ (Au+Au) and 2400 (Pb+Pb).
For the transverse density {\em profile}, on the other hand,  
we use the same binary collision distributions as for jets.
This choice is motivated by a classical Yang-Mills 
(``color glass condensate'') calculation\cite{CYMecc}
that found eccentricities at early times to be much closer
to binary collision eccentricities than to wounded nucleon
 eccentricities.

The initial conditions above are more general than they may appear
at first sight. Due to scalings\cite{nonequil} 
of the transport equation (\ref{Eq:BTE}), 
a single transport
solution contains the answer to a whole class of {\em equivalent} problems
with scaled initial conditions, cross sections, and particle properties. 
For example, one can freely 
increase the initial density provided cross sections
are reduced in inverse proportion to keep the mean free path the same. 
The initial 
temperature can also be rescaled together with particle masses
(in fact for our massless quanta, $T_0$
does not influence the density evolution at all).

Though in principle the transport provides an ensemble of evolving 
scattering centers, here 
we solely use density information to calculate
energy loss. Technically this is very similar to employing 
ideal or viscous hydrodynamics for the medium evolution. 
We plan to explore different bulk dynamics models in the future.

\section{Results}
\label{Sc:Results}

Below we 
focus on two basic high-$p_T$ observables for neutral pions at 
midrapidity, 
the nuclear suppression factor 
$R_{AA}$ and the momentum anisotropy (elliptic flow)
$v_2 = \langle \cos 2\phi \rangle_{p_T}$.
Only energy loss is considered, i.e.,
contributions by the radiated gluons to the final spectrum
and feedback on the bulk medium due to the jet were ignored.
This is a good approximation at sufficiently high $p_T$.

We consider {\bf four scenarios}
based on i) whether the medium is only undergoing Bjorken expansion 
(``1D'' as in \cite{CUJET1p0,Horowitzv2}) 
or transverse expansion as well (``3D'');
and ii) whether average energy loss $\langle\Delta E\rangle$ 
is used or the stochastic $\Delta E(z)$.

\begin{figure}[h]
\epsfysize=6.15cm
\epsfbox{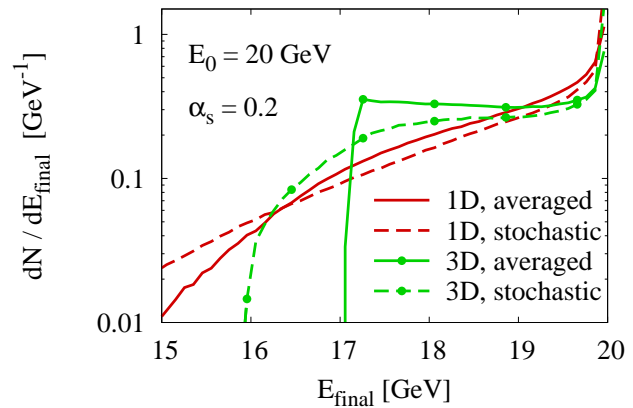}
\caption{Final energy distribution for gluon jets of fixed initial energy
20~GeV in four scenarios (see text) for $Au+Au$ at 
$\sqrt{s_{NN}}=200$~GeV with $b=8$~fm,
using averaged or stochastic energy (solid vs dashed) 
with or without transverse expansion for the medium (lines without symbols vs
lines with symbols). All distributions are normalized to unit area.
In the 3D case, the medium evolution was computed using the parton
transport MPC\cite{MPC}.}
\label{Fig:dNdEloss_20GeV}
\end{figure}

Before we turn to results for RHIC and LHC, it is illustrative to compare 
these scenarios for  gluon jets of {\em fixed} initial energy in 
$Au+Au$ at RHIC
with $b=8$~fm. Every aspect of the calculation
(such as 
production points with binary collision profile, uniform angles in azimuth,
etc)
is as discussed above except that the gluon jets all start out with energy 
$E_0 = 20$~GeV. 
Figure~\ref{Fig:dNdEloss_20GeV} shows the final jet energy 
distribution after energy loss.
The first obvious feature is that in both 1D and 3D scenarios, 
stochastic energy loss gives a broader energy loss distribution
than using the average $\langle E\rangle$. This is natural since
averaging reduces fluctuations in general. Given that
parton spectra are convex at high $p_T$, more fluctuation with stochastic
energy loss should result in weaker nuclear suppression (higher $R_{AA}$),
for the same $\alpha_s$.

Another noticeable feature in Fig.~\ref{Fig:dNdEloss_20GeV} 
is that the distribution is 
{\em narrower} with transverse expansion (3D) 
than with longitudinal expansion only (1D). This is a result of the
 interplay between energy loss and transverse expansion of the density profile.
In the ``frozen'' 1D case, 
largest energy loss occurs for jets that start spatially
out of plane and move through the center of the collision zone along 
the long axis of the almond-shaped
transverse profile. With transverse expansion, however, the situation is
the opposite. Largest energy loss occurs for jets that start in-plane and 
move along the {\em initially} shorter axis of the almond. Because the medium
expands faster in the in-plane direction, 
by the time these jets cross the center, the spatial eccentricity
largely disappears (near the collision 
center the elongation is actually in-plane
already but the tail of the density distribution away from the center still 
keeps
traditional eccentricity 
$\varepsilon = \langle y^2 - x^2\rangle / \langle x^2 + y^2\rangle$ positive).
The low-energy side of the final jet energy distribution 
in Fig.~\ref{Fig:dNdEloss_20GeV} cuts off 
more quickly in the 3D case because
the probability for hard scattering (forming a jet) 
at the edge of the collision zone drops
much sharper in-plane than out-of-plane. 

Rapid transformation of the collision zone towards a more axially symmetric
shape
also reduces the energy loss 
difference between
jets crossing through the collision center in-plane vs
out-of plane (what in-plane jets miss in opacity while approaching the center 
they make up for while leaving it, and vice versa), which explains 
why the energy loss distribution is much flatter in the transversely expanding
case.

\subsection{Nuclear suppression in Au+Au at RHIC}

\begin{figure}[h]
\epsfysize=6.15cm
\epsfbox{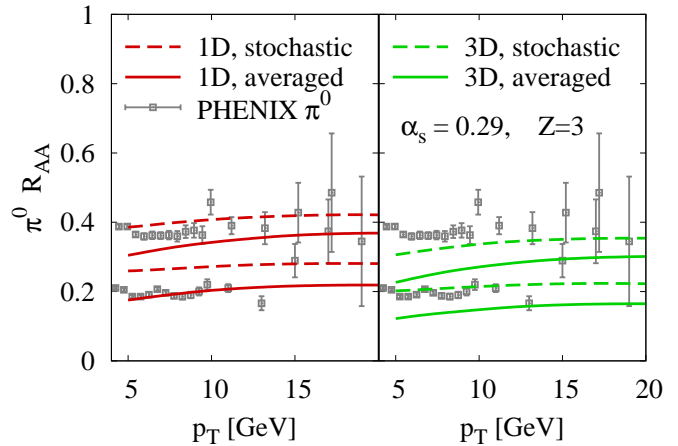}
\caption{Neutral pion $R_{AA}$ in $Au+Au$ at $\sqrt{s_{NN}}=200$~GeV at RHIC
with impact parameters $b=3$ fm and $8$~fm 
(lower pair of lines vs upper pair of lines, respectively, in both panels) 
calculated in four different scenarios with GLV energy loss (see text).
{\em Left panel:} transversely frozen density profiles. {\em Right panel:} 
realistic transverse expansion modeled with parton transport MPC\cite{MPC}.
{\em Dashed curves} were obtained with fluctuating energy loss, while 
{\em solid curves}
with averaged energy loss along the jet path.
Data\cite{PHENIX_pi0_y2008} from PHENIX (boxes) are also shown for comparable
centralities 0-10\% and $\approx 30$\% to guide the eye.}
\label{Fig:RHIC_RAA}
\end{figure}

Figure~\ref{Fig:RHIC_RAA} shows our results for neutral pion 
$R_{AA}$ at RHIC, for fixed
$\alpha_s = 0.29$ for all four scenarios.
With stochastic energy loss the suppression is noticeably weaker, 
as expected for ``upward curving'' parton spectra at high $p_T$. 
Quite interestingly, realistic transverse
expansion significantly enhances jet quenching.
This is a generic feature of GLV energy loss
coming from the $(1-\cos)$
interference term in (\ref{GLV1}). Scatterings at large $z$
induce larger energy loss (cf. Fig.~\ref{Fig:I}), 
and with a transversely expanding 
density profile there is higher chance to scatter further away from the 
production point than in the transversely frozen case.

\begin{figure}[h]
\epsfysize=6.15cm
\epsfbox{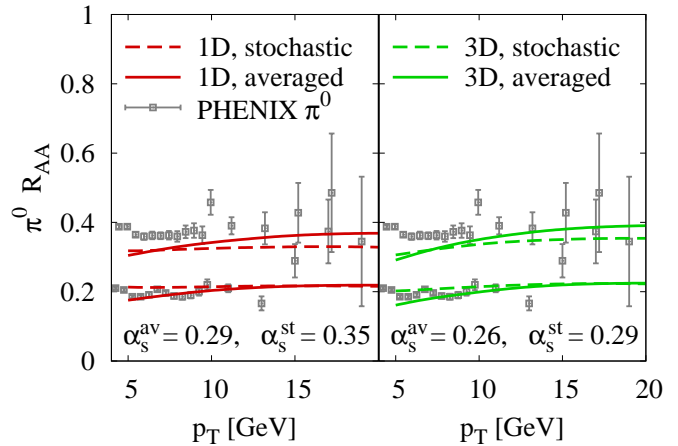}
\caption{Same as Fig.~\ref{Fig:RHIC_RAA} but with $\alpha_s$ tuned 
in all four scenarios to reproduce $R_{AA}$ in nearly central ($b=3$ fm) 
collisions.}
\label{Fig:RHIC_RAA_tuned}
\end{figure}

Unfortunately,
without precise control over $\alpha_s$,
$R_{AA}$ alone cannot differentiate between our four scenarios.
As shown in Fig.~\ref{Fig:RHIC_RAA_tuned}, 
after a slight tuning of $\alpha_s$ to reproduce the
suppression in central collisions, differences in $R_{AA}$ 
largely disappear. We need $\sim 10-20$\% higher $\alpha_s$ with fluctuating 
energy loss than with the path-averaged one, 
while $\sim 10-15$\% 
lower $\alpha_s$ with transverse expansion than without it.

\subsection{Elliptic flow in Au+Au at RHIC}

\begin{figure}[h]
\epsfysize=6.15cm
\epsfbox{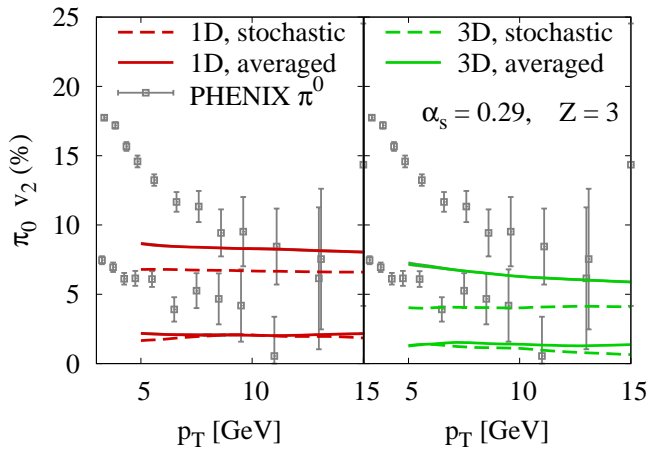}
\caption{Neutral pion $v_2$ in $Au+Au$ at $\sqrt{s_{NN}}=200$~GeV at RHIC
with impact parameters $b=3$ fm and $8$~fm 
(lower pair of lines vs upper pair of lines, respectively, in both panels) 
calculated in four different scenarios with GLV energy loss (see text).
{\em Left panel:} transversely frozen density profiles. {\em Right panel:} 
realistic transverse expansion modeled with parton transport MPC\cite{MPC}.
{\em Dashed curves} were obtained with fluctuating energy loss, while 
{\em solid curves}
with averaged energy loss along the jet path.
Data\cite{PHENIX_pi0_v2_y2010} from PHENIX (boxes) 
are also shown for comparable
centralities 0-10\% and $\approx 30$\% to guide the eye.}
\label{Fig:RHIC_v2}
\end{figure}

Luckily, $p_T$-differential elliptic flow turns out to be
a much more
sensitive probe of the medium evolution and energy loss treatment.
Figure~\ref{Fig:RHIC_v2} shows our results
for neutral pion 
$v_2(p_T)$ in $Au+Au$ at $\sqrt{s_{NN}}=200$~GeV at RHIC, 
for fixed $\alpha_s = 0.29$ in all four scenarios. We find that, 
especially in more peripheral $b=8$~fm collisions, stochastic energy loss
gives smaller elliptic flow both with and without transverse expansion
than average energy loss.
This is in line with the weaker
nuclear suppression for the stochastic case shown
in Fig.~\ref{Fig:RHIC_RAA}. 

On the other hand,
transverse expansion {\em reduces} elliptic flow, and at the same time
also gives {\em smaller} $R_{AA}$ as we have seen in the previous 
Section. This
may seem counter-intuitive at first, but it is also a
feature of GLV energy loss. Because the
interference term in (\ref{GLV1})
biases against early scattering in the medium, a jet parton loses
more energy if it scatters further away from the production point,
which also means the scattering is later in time.
Compared to the transversely ``frozen'' 1D
scenario, transverse expansion gives at later times 
higher densities away from the center, so there is more quenching, 
but at the same time the medium 
also rapidly becomes more azimuthally symmetric, so the elliptic flow response
is weaker.

\begin{figure}[h]
\epsfysize=6.15cm
\epsfbox{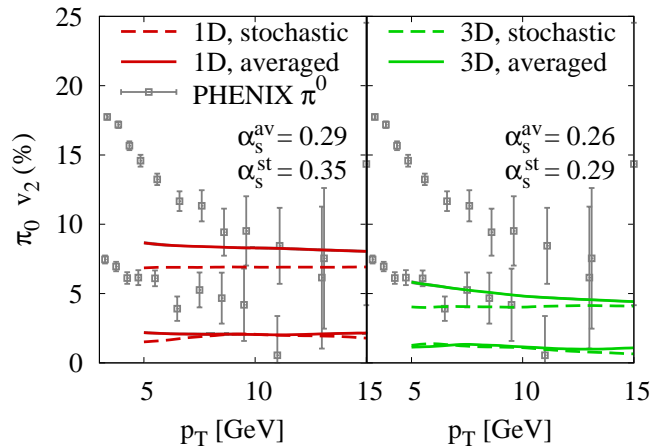}
\caption{Same as Fig.~\ref{Fig:RHIC_v2} but with $\alpha_s$ tuned 
in all four scenarios to reproduce $R_{AA}$ in nearly central ($b=3$ fm) 
collisions.}
\label{Fig:RHIC_v2_scaled}
\end{figure}

The reduction of $v_2$ in the 3D case 
is manifest even after $\alpha_s$ is tuned to reproduce $R_{AA}$ 
in central collision, as shown in Fig.~\ref{Fig:RHIC_v2_scaled}.
For both averaged energy and stochastic energy loss, 
$v_2$ is reduced by $\approx 40$\% (nearly half!) at high 
$p_T\sim 10-15$~GeV 
due to transverse
expansion. Similar reduction is present at LHC energies as
we discuss in Section~\ref{Sc:LHC_v2}.
In the more realistic ``3D'' scenario 
with both transverse and longitudinal expansion,
both averaged and stochastic energy loss give about
the same $v_2(p_T)$ at high $p_T$ for both centralities we studied.

\subsection{Nuclear suppression in Pb+Pb at the LHC}

Now we turn to results for Pb+Pb at LHC energies. 
For the nuclear suppression,
we find the same generic features as in Au+Au at RHIC: i) medium evolution
with transverse expansion
gives stronger suppression (smaller $R_{AA}$) than the transversely ``frozen''
scenario,  and ii) stochastic energy
loss gives less suppression than the path-averaged 
$\langle \Delta E\rangle$.

\begin{figure}[h]
\epsfysize=6.15cm
\epsfbox{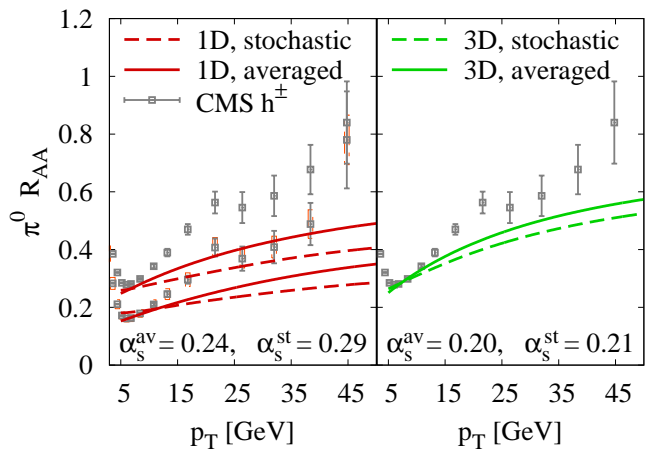}
\caption{Neutral pion $R_{AA}$ in $Pb+Pb$ at $\sqrt{s_{NN}}=2.76$~TeV 
at the LHC
with impact parameters $b=3$ fm (left panel only) and $8$~fm (both panels) 
calculated in four different scenarios with GLV energy loss (see text).
{\em Left panel:} transversely frozen density profiles. {\em Right panel:} 
realistic transverse expansion modeled with parton transport MPC\cite{MPC}.
{\em Dashed curves} were obtained with fluctuating energy loss, while 
{\em solid curves}
with averaged energy loss along the jet path.
Charged hadron data\cite{CMS_ch_y2012} from CMS (boxes) 
are also shown, scaled to comparable
centralities 0-10\% and $\approx 30$\%, to guide the eye.}
\label{Fig:LHC_RAA_tuned}
\end{figure}

Figure~\ref{Fig:LHC_RAA_tuned} shows our results for neutral pion 
$R_{AA}$ in Pb+Pb at $\sqrt{s_{NN}}=2.76$~TeV at the LHC, 
{\em after} $\alpha_s$ has been adjusted to reproduce 
the suppression measured around $p_T \sim 8$~GeV. 
At LHC energies
we only computed the transport evolution for $b=8$~fm collisions,
therefore in the 3D scenarios we adjust $\alpha_s$ to $R_{AA}$ measurements
for $\approx 30$\% centrality($b=8$~fm) instead of $0-10$\% central
collisions ($b\approx 3$~fm).
We generally find that in all cases studied 
stochastic energy loss gives somewhat
flatter $R_{AA}$ than a computation using the average energy loss. 
From our fixed-coupling calculation, $R_{AA}$ 
curves rise with $p_T$ slower than the trend in the data.
Incorporation of running $\alpha_s(Q^2)$,
as was done, e.g., in~\cite{CUJETrunning},
should yield better agreement.

We need somewhat larger $\alpha_s$ values to match the observed suppression
at RHIC than to reproduce suppression at the LHC
(Fig.~\ref{Fig:RHIC_RAA_tuned} vs Fig.~\ref{Fig:LHC_RAA_tuned}),
i.e., the medium at the LHC appears somewhat less opaque to jets 
than extrapolation
based on RHIC data would suggest. 
Our results thus qualitatively reinforce the findings 
in~\cite{LHCvsRHICcoupling},
although we see about twice as large an effect.
In our transversely static (1D) scenarios
the ratio of LHC to RHIC effective couplings is
about $\alpha_s^{LHC}/\alpha_s^{RHIC} \approx 0.8$,
while with transversely expanding medium
the ratio is somewhat
lower, $\alpha_s^{LHC}/\alpha_s^{RHIC} \sim 0.75$.

\subsection{Elliptic flow in Pb+Pb at the LHC}
\label{Sc:LHC_v2}

\begin{figure}[h]
\epsfysize=6.15cm
\epsfbox{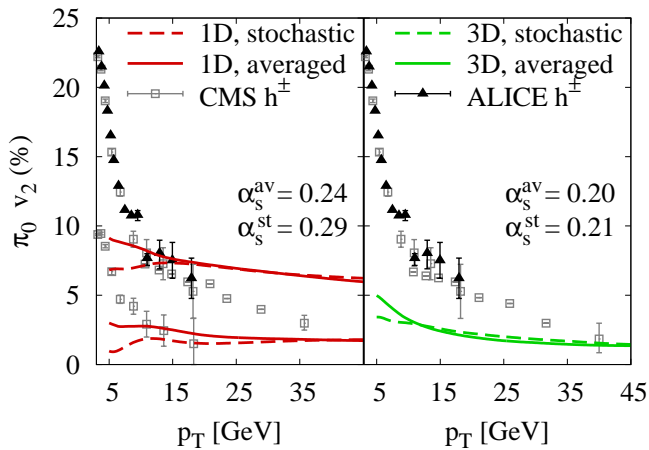}
\caption{Neutral pion $v_2(p_T)$ in $Pb+Pb$ at $\sqrt{s_{NN}}=2.76$~TeV 
at the LHC
with impact parameters $b=3$ fm (left panel only) and $8$~fm (both panels) 
calculated in four different scenarios with GLV energy loss (see text).
{\em Left panel:} transversely frozen density profiles. {\em Right panel:} 
realistic transverse expansion modeled with parton transport MPC\cite{MPC}.
{\em Dashed curves} were obtained with fluctuating energy loss, while 
{\em solid curves}
with averaged energy loss along the jet path.
Charged hadron data\cite{CMS_v2_ch_y2012,ALICE_v2_ch_y2012} from CMS (boxes) 
and ALICE (triangles)
are also shown, scaled to comparable
centralities 0-10\% and $\approx 30$\%, to guide the eye.}
\label{Fig:LHC_v2_tuned}
\end{figure}

Finally we demonstrate that with GLV energy loss the
striking reduction of high-$p_T$ 
elliptic flow seen at RHIC energies for transversely expanding media
also occurs at LHC energies.
Figure~\ref{Fig:LHC_v2_tuned}
shows our results
for neutral pion 
$v_2(p_T)$ in $Pb+Pb$ at $\sqrt{s_{NN}}=2.76$~TeV at the LHC,
again after $\alpha_s$ has been adjusted to reproduce 
$R_{AA}(p_T \sim 8~{\rm GeV})$. We only find a very small reduction of elliptic
flow due to stochastic energy loss compared to using the average $\langle \Delta E\rangle$ along the jet path. The difference between transversely ``frozen'' evolution and full 3D expansion,
however, is even larger than in Fig.~\ref{Fig:RHIC_v2_scaled} for 
RHIC collisions. 
By $p_T = 5$~GeV, $v_2$ is reduced by more than half, 
and above $p_T = 15$~GeV by more than two-thirds,
irrespectively of whether we use average energy loss 
or the stochastic one.

\section{Conclusions}

In this work we investigated nuclear suppression and elliptic flow
in $A+A$ reactions at RHIC and the LHC 
using Gyulassy-Levai-Vitev (GLV) energy loss
with the covariant transport model MPC\cite{MPC} for the bulk medium.
We compared calculations with transversely ``frozen'' density profiles
(as in~\cite{CUJET1p0,Horowitzv2})
to calculations with realistic transverse expansion,
and also studied the difference between 
path-averaged (deterministic) energy
loss and stochastic energy loss that fluctuates depending on the location
of the scattering center that interacts with the jet parton.

Our most striking finding is
that, at both RHIC and LHC, realistic transverse expansion strongly suppresses 
elliptic flow at high $p_T$ compared to calculations with longitudinal
Bjorken expansion only. We argue that this is a generic
feature of GLV energy loss, coming from interference
bias against early rescattering in the medium.

We also find that transverse expansion enhances high-$p_T$ suppression, 
while energy loss fluctuations in the stochastic case lead to weaker
suppression and smaller elliptic flow.
However, unlike the strong 
reduction of elliptic flow with transverse expansion,
these latter effects nearly disappear once calculations are 
adjusted to reproduce $R_{AA}$ in central collisions.

Though our calculation lacks a few relevant effects
such as elastic energy loss and multi-gluon emission,
the results do suggest that GLV energy loss will have difficulty
with the simultaneous description of nuclear suppression and elliptic flow
at RHIC and the LHC.
In view of the $R_{AA}(\phi)$ challenge posed 
for perturbative QCD energy loss in a recent 
compilation\cite{PHENIXeloss} by PHENIX,
we believe it is imperative to test our findings with more full-fledged
GLV implementations\cite{CUJET1p0,Horowitzv2},
and also other bulk evolution models (such as hydrodynamics).

\begin{acknowledgments}
We thank A.~Buzzatti, I.~Vitev, M.~Gyulassy and W.~Horowitz for stimulating 
discussions. 
This work was supported by the US DOE under grant
DE-PS02-09ER41665. D.S. was partially supported by the JET Collaboration
(DOE grant DE-AC02-05CH11231).
\end{acknowledgments}

\appendix

\section{The energy loss integral $I$}
\label{App:I}

Here we discuss how Eq.~(\ref{Eq:deltaEz}) is obtained. Substituting
(\ref{GLV1}) and (\ref{kin_bounds}) 
into (\ref{Eq:deltaEz_gen}) leaves one with five integrals to do.
Axial symmetry $d^2 {\bf q}\, d^2 {\bf k}\, dx
 = (\pi/2)\, d(q^2)\, d(k^2)\, d\varphi\, dx$  reduces these to four,
and with dimensionless
variables 
\be
\kappa = \frac{k}{\mu} \ , \quad 
\xi = \frac{q}{\mu} \ , \quad
\epsilon = \frac{E}{\mu} \ , \quad
b = \frac{z}{\tau} \ ,
\ee
and $u = ({\bf k} - {\bf q})^2 /\mu^2 = 
\kappa^2 + \xi^2 - 2\kappa \xi \cos\varphi$, we have
\bea
I(b, \epsilon) &=&
            \int\limits_0^{3\epsilon} d(\xi^2)
            \int\limits_0^{\pi} \frac{d\varphi}{\pi}
            \int\limits_0^{\epsilon} d\kappa
            \int\limits_{x_{min}}^1 dx \,
           \frac{2 \xi\,\cos\varphi}{(\xi^2 + 1)^2 u}
\nonumber \\
&&\qquad\qquad\, \times
\left( 1 -   \cos \frac{u\, b}{x} \right)
\label{Eq:I}
\eea
where the lower limit on the $x$ integral is 
$x_{\min} = \max(1, \kappa) / \epsilon$. 
The innermost integral is doable with the help of
\be
\int dx\, \cos \frac{\alpha}{x} = 
x \cos \frac{\alpha}{x} 
+ \alpha\, {\rm Si} \frac{\alpha}{x}
\ee
where $\rm Si$ is the sine integral function.
The remaining three integrals can be evaluated numerically (at some
expense because the integrand is oscillatory).

For a numerically easier approximate result, one may momentarily
ignore the upper bound (\ref{kin_bounds}) on $q$. With new variable 
${\bf q} \to {\bf Q} = ({\bf k} - {\bf q})/\mu$, ${\bf Q}$ is then unrestricted
and we have, 
again in polar coordinates $d^2 {\bf Q}\, d^2\, {\bf k}\, dx  = (\pi/2)\, 
d(Q^2)\, d(k^2)\, d\phi\, dx$,
\bea
I(b, \epsilon) &\to &
            \int\limits_{1/\epsilon}^1 dx 
            \int\limits_0^\infty \frac{du}{u}
\left( 1 -   \cos \frac{u\,b}{x} \right)
\nonumber \\
&&\times           \int\limits_0^{x \epsilon} d\kappa
            \int\limits_0^{\pi} \frac{d\phi}{\pi}
           \frac{2(\kappa - Q\cos\phi)}
                {(\kappa^2 + u - 2\kappa Q \cos\phi + 1)^2} \ .
\nonumber\\
\eea
The $\phi$ and $\kappa$ integrals evaluate to
\be
g(u,x\epsilon) = 
\frac{1}{1+u} - \frac{1}{\sqrt{x^2 \epsilon^2 + 2x \epsilon (1-u)+(1+u)^2}}
\ee
Restoring now an approximate $\phi$-averaged 
upper limit on $Q^2$, $Q^2 = u \sim \kappa^2 + \xi^2 
\le x^2 \epsilon^2 + 3\epsilon < 4x \epsilon^2$ gives
the simpler 2D integral
\be
I_{2D}(b, \epsilon) =
            \int\limits_{1/\epsilon}^1 dx 
            \int\limits_0^{4x\epsilon^2} \frac{du}{u}
\left( 1 -   \cos \frac{u\,b}{x} \right) g(u,x\epsilon) \ ,
\label{Eq:I_2D}
\ee
which approximates the full result (\ref{Eq:I}) well in practice 
provided $E/\mu \gton 10$ (cf. ~Fig.~\ref{Fig:I}).


\end{document}